\documentclass[%
reprint,
superscriptaddress,
amsmath,amssymb,
aps,
prl
]{revtex4-2}

\usepackage{graphicx}% Include figure files
\graphicspath{ {Figures/} }

\usepackage{dcolumn}% Align table columns on decimal point
\usepackage{bm}% bold math
\usepackage[export]{adjustbox}
\usepackage[normalem]{ulem}

\usepackage{mathtools} % Braket notation
\DeclarePairedDelimiter\bra{\langle}{\rvert}
\DeclarePairedDelimiter\ket{\lvert}{\rangle}
\DeclarePairedDelimiterX\braket[2]{\langle}{\rangle}{#1 \delimsize\vert #2}

\usepackage[utf8]{inputenc}
\usepackage[T1]{fontenc}

\usepackage{mathrsfs}
\usepackage{dsfont}

\usepackage{amssymb}
\usepackage{amsmath}
\usepackage{amsfonts}
\usepackage{amsthm}

\usepackage{hyperref}
\hypersetup{pdfpagemode=UseNone}

\newcounter{rem}
\newcommand{\mc}[1]{\mathcal{#1}}

\DeclarePairedDelimiter\ceil{\lceil}{\rceil}

\def\>{\rangle}
\def\<{\langle}
\newcommand{\proj}[1]{| #1 \rangle\! \langle #1 |}
\newcommand{\idty}{\mathds{1}}
\def\tr{{\rm tr}}
\def\pr{{\rm Pr}}

\def\rho{{\varrho}}

\def\textbf#1{{\bf #1}}
\newcommand{\Cx}{\mathbb{C}}

\newcommand{\Rl}{\mathds{R}}
\gdef\lvert{\delimiter"426A30C }
\gdef\rvert{\delimiter"526A30C }
\gdef\lVert{\delimiter"426B30D }
\gdef\rVert{\delimiter"526B30D }
\newcommand{\norm}[1]{\lVert#1\rVert}

\usepackage{xcolor}

\begin{document}

%\title{Canonical typicality for generalized subsystems}

\title{Canonical typicality under general quantum channels}

\author{Pedro S. Correia}
\affiliation{Departamento de Ciências Exatas, Universidade Estadual de Santa Cruz, Ilhéus, Bahia 45662-900, Brazil}
\email{pscorreia@uesc.br}
\author{Gabriel Dias Carvalho}
\affiliation{Física de Materiais, Universidade de Pernambuco, 50720-001, Recife, PE, Brazil}
\affiliation{Instituto de Física, Universidade Federal Fluminense, Av. Litoranea s/n, Gragoatá 24210-346, Niterói, RJ, Brazil}
\author{Thiago R. de Oliveira}
\affiliation{Instituto de Física, Universidade Federal Fluminense, Av. Litoranea s/n, Gragoatá 24210-346, Niterói, RJ, Brazil}
\author{Ra\'ul O. Vallejos}
\author{Fernando de Melo}
\affiliation{Centro Brasileiro de Pesquisas F\'{\i}sicas, Rua Dr. Xavier Sigaud, 150, Rio de Janeiro, RJ, Brazil}
\date{\today}

\begin{abstract}
With the control of increasingly complex quantum systems, the relevant degrees of freedom we are interested in may not be those traditionally addressed by statistical quantum mechanics. Here we employ quantum channels to define generalized subsystems, capturing the pertinent degrees of freedom, and obtain their associated canonical state. We show that the generalized subsystem description from almost any microscopic pure state of the whole system will behave similarly to its corresponding canonical state. Such canonical typicality behavior depends on the entropy of the channel used to define the generalized subsystem.
\end{abstract}
\maketitle

\textit{Introduction.} Ensembles are key ingredients in pillar fields as thermodynamics and statistical mechanics. An ensemble formalizes the idea that for complex systems one cannot control all its degrees of freedom as to prepare it in  a well-defined state. In each run of the experiment only few ``macroscopic'' quantities are fixed, and any ``microscopic'' state that is consistent with the macroscopic quantities can be prepared. The collection of such compatible microscopic states forms the ensemble of a given experimental scenario.

Two of the most prevailing ensembles in physics are the microcanonical and the canonical ensembles. For the first the system is isolated and formed by a large number of degrees of freedom, with some macroscopic properties having well defined values. The second can be seen as the description of a subsystem, with fixed number of degrees of freedom, of the microcanonical ensemble. The canonical ensemble thus characterizes the situation where  macroscopic properties are allowed to vary within a given subsystem, while  such properties are fixed when considering the whole system.

While the use of ensembles has been very successful, they introduce a probabilistic character to the microscopic description, which can be argued deterministic: In a given run of an experiment one might expect that all properties of the probed system are fixed, and not just the few controlled macroscopic quantities. Within this reasoning, the microscopic state is well defined, albeit unknown.  To reconcile this deterministic perspective with the overwhelming success of the probabilistic description is one of the main discussions in the foundations of statistical mechanics since Boltzmann. The most common justification, using the microscopic laws of mechanics, invokes chaos theory and the ergodic hypothesis~\cite{lebowitz1993boltzmann}. Others consider an information theory approach, exploiting the principle of maximum entropy~\cite{jaynes106,jaynes108}. 

Another possibility is the so-called typicality argument: almost all microscopic states which are compatible with a given preparation procedure will behave similarly to the ensemble description for any macroscopic property~\cite{lebowitz1993boltzmann}. Such a method had been already proposed by Boltzmann, and found its way to the quantum domain via different approaches to obtain coarse-grained statistics from quantum systems~\cite{neumann29,Farquhar_57, Bocchieri_58,Zwanzig}. The latter has remained forgotten until the recent reanalysis of the typicality argument from the quantum information perspective~\cite{popescu2006, reimann2008,gemmer_book}  -- finding applications in areas ranging from entanglement theory~\cite{hayden2006aspects, Tiersch_2013}, quantum thermodynamics~\cite{berg_2023}, passing through condensed matter~\cite{SpinCurrent}, and going all the way to black-hole theory~\cite{lloydPhd, hayden2007black}. In this modern mindset, \textit{canonical typicality} asserts that any pure state from the whole system, that abides by the fixed macroscopic quantities defining a preparation scheme, will be locally, i.e., for the subsystem, close to the canonical state. How close in this perspective will depend on various ingredients and are somehow a formalization of the ``thermodynamical limit'' (details below).

Crucial in the above discussion is the concept of a subsystem with a fixed number of degrees of freedom. Traditionally we equate this with the idea of a fixed number of particles, which is a very natural attitude when dealing, for instance, with a gas of weakly interacting atoms. More recently the idea of subsystems was extended to cover collective degrees of freedom. An example of that is a superconducting metal, with Cooper pairs forming individual entities which are delocalized in space: here the quasi-particle description is more pertinent~\cite{superconductivity}. With the control of complex quantum systems new forms of subsystems are becoming relevant.  Think for instance of an optical lattice where individual atoms are loaded onto potential wells~\cite{fukuhara, gross2017quantum}, but their reading is done in blocks that contains multiple wells: in this case a coarse-grained description is more convenient~\cite{cris2017,pedrinho}.

To deal with all possible definitions of subsystems, we employ the theory of quantum channels, following~\cite{alicki2009}, to define generalized subsystems, and we put forward their corresponding canonical ensemble. We also establish the conditions for these generalized scenarios to admit a canonical typicality reasoning, thus extending the typicality interpretation to the foundations of statistical mechanics over vast and meaningful situations. The ``thermodynamics'' of a given experimental scenario thus depends on what can be measured, i.e., on what can be called an effective particle.

\noindent \textit{Canonical typicality.} We briefly review the basics of canonical typicality for the case in which the subsystem is just a subset of the total system~\cite{popescu2006,goldstein2006}. Consider a total system to which we assign a Hilbert space $\mc{H}_T$ -- for instance, a tensor product of $N$ individual particles' spaces. To define a subsystem we split the total space in two parts, $\mc{H}_T= \mc{H}_S\otimes\mc{H}_E$; with $\mc{H}_S$ associated with the subsystem, and $\mc{H}_E$ associated with the rest of the system and it is seen as an environment for the first. We then define a restricted subspace $\mc{H}_R\subseteq \mc{H}_S\otimes \mc{H}_E$  in which all the states abide by the macroscopic constraints. 

The total system's description, which represents ignorance of any other constraint than the already contemplated by the restriction, is the microcanonical state $\mc{E}_R=\idty_R/d_R$, with $\idty_R$ the projector into $\mc{H}_R$ and $d_R=\dim(\mc{H}_R)$. The system's canonical state $\Omega_{\tr_E}$ is then obtained by  tracing out the environmental degrees of freedom: $\Omega_{\tr_E}=\tr_E(\mc{E}_R)$. 

The canonical typicality approach is based in showing that for almost all states $\ket{\psi}\in\mc{H}_R$, taken uniformly at random (from the unitarily invariant measure), the local state of the subsystem, $\rho_{\tr_E}^\psi=\tr_E(\proj{\psi})$, is close to the canonical state $\Omega_{\tr_E}$ when the effective dimension of the environment, $d_E^\text{eff}=1/\tr(\Omega_{\tr_S}^2)$ (with $\Omega_{\tr_S}=\tr_S(\mc{E}_R)$), is much bigger than the subsystem dimension $d_S=\dim(\mc{H}_S)$. It was shown that the average distance is bounded as follows:
\begin{align}
\overline{\mc{D}(\rho_{\tr_E}^\psi,\Omega_{\tr_E})}^\psi \le \frac{1}{2}\sqrt{\frac{d_S}{d_E^\text{eff}}}.
\label{eq:popescubound}
\end{align}
Hereafter, the overbar means the average taken over the uniform measure, and $\mc{D}(\rho,\sigma)=\norm{\rho-\sigma}_1/2$ with the trace norm defined as $\norm{A}_1=\tr\sqrt{A^\dagger A}$. Moreover, employing  Levy's lemma~\cite{levy,MarkusThesis} (see Section I in Supplemental Material), they showed that the probability for $\rho_{\tr_E}^\psi$ being further away from $\Omega_{\tr_E}$  decays exponentially with $d_R$. The canonical state is then recovered from any pure state in the restricted subspace when $d_R\gg 1$ and $d_S\ll d_E^\text{eff}$, which are then the formal requirements for equilibration.

\noindent \textit{Generalized subsystems.} In the above scenario, the subsystem is a simple partition of the pre-established total Hilbert space. Within statistical mechanics it is commonsensical to split the total system into a system of interest, the intended subsystem, and its environment, for which no control is assumed. Such a split is often possible due to the weak interaction between the two parts.  By partial tracing the environmental degrees of freedom, we obtain the subsystem description.

However, to find a physically motivated partition some reshape the total Hilbert space might be necessary. A text-book example of this is the Hydrogen atom case. A initial description in terms of an electron and a proton is rearranged into the degrees of freedom of center-of-mass and relative-particle. In this new split,  both effective particles are decoupled and we can solve Schrödinger's equation. The idea of quasiparticles, ubiquitous in condensed matter physics, is a somewhat more recent example of this reshuffling of Hilbert' space in order to find meaningful effective descriptions. Once the more compelling degrees of freedom are found, the remaining ones can be seen, and treated, as an effective environment.

In the aforementioned cases, the Hilbert space rearrangement was done by a unitary transformation (like Bogoliubov transformations~\cite{bogoliubov}), and the focus on a subsystem was done by tracing out the weakly interacting degrees of freedom of the effective environment. In the quantum information lingo, this definition of subsystem is specified by the quantum channel $\Lambda_{\tr_E\circ U}:\mc{L}(\mc{H}_T)\mapsto \mc{L}(\mc{H}_S)$ which acts as $\Lambda_{\tr_E}^U(\rho)=\tr_E(U \rho U^\dagger)$. The canonical typicality remains accurate for this unitary reshuffle.

While very useful and successful, this is not the only way to obtain effective subsystems of quantum systems. As established in~\cite{alicki2009}, and further developed in~\cite{cris2017,pedrinho,oleg,gabriel2020,cris2020,oleg2020a,carlospineda2021,correia2021,vallejos2022}, generalized subsystems can be achieved by exploiting the most general form of a quantum channel $\Lambda:\mc{L}(\mc{H}_T)\mapsto\mc{L}(\mc{H}_S)$~\cite{watrous}:
 \begin{equation}
 \Lambda(\rho)=\tr_N (V \rho V^\dagger).
 \label{eq:channel}
 \end{equation}
 In the above expression $V:\mc{H}_T\mapsto \mc{H}_S\otimes\mc{H}_N$ is an isometry. Like unitary transformations, isometries also preserve the scalar product as $V^\dagger V = \idty_T$. However, as isometries might change the space dimension, $V V^\dagger = \idty_{SN}$ is a projector onto $\mc{H}_S\otimes\mc{H}_N$. It is this very change of dimensions, empowered by the isometry, which  can be exploited to obtain generalized subsystems. $\mc{H}_T$ is reshuffled to $\mc{H}_S\otimes\mc{H}_N$, with $\mc{H}_N$ being associated with an effective (possibly mathematically abstract) environment for the generalized subsystem acting on $\mc{H}_S$. Like before, the now general effective environment is discarded by the partial trace. An explicit example of a generalized subsystem will be given in what follows.

\noindent \textit{Canonical typicality for generalized subsystems.} Here we present the main result of this work.  Consider a system to which we assign a total space $\mc{H}_T$, and suppose that such a system obeys some arbitrary restriction $R$, and thus act on $\mc{H}_R\subseteq \mc{H}_T$. Like before, the  description of the system  is given by its microcanonical state $\mathcal{E}_R=\mathds{1}_{R}/d_R$. Our generalized canonical state, i.e., the description of the generalized subsystem, will be given by the action of a general quantum channel $\Lambda: \mathcal{L}(\mathcal{H}_R)\mapsto\mathcal{L}(\mathcal{H}_S)$ on the microcanonical state: $\Omega_\Lambda:=\Lambda(\mc{E}_R)$. 

For the generalized canonical typicality principle we are then interested in showing that for almost any $\ket{\psi}\in \mc{H}_R$, taken uniformly at random,  we will have $\rho_\Lambda^\psi:=\Lambda(\proj{\psi}) \sim \Omega_\Lambda$. Following Ref.\cite{popescu2006}, this is done in two parts. 

The first part is to bound the average distance between $\rho_\Lambda^\psi$ and $\Omega_\Lambda$. In Section II of Supplemental Material we show that 
\begin{equation}
\overline{\mathcal{D}(\rho_\Lambda^\psi,\Omega_\Lambda)}^\psi\leq\dfrac{1}{2}\sqrt{d_S\tr({J_\Lambda}^2)} =\dfrac{1}{2}\sqrt{d_S(1-S_L(\Lambda))}.
\label{eq:ineqentropymain}
\end{equation}
Here $J_\Lambda:=\Lambda\otimes\idty(\proj{\phi^+})$ is the Choi state of the channel $\Lambda$, with $\ket{\phi^+}=\sum_i\ket{ii}/\sqrt{d_R}\in\mc{H}_R\otimes\mc{H}_R$.  The Choi state $J_\Lambda$ is isomorphic to $\Lambda$. Using this relationship, in~\cite{roga10,roga11}  $\tr(J_\Lambda^2)$ was defined as the channel purity, and $S_L(\Lambda)=1-\tr(J_\Lambda^2)$ as the channel linear entropy.

As expected, the bound in \eqref{eq:ineqentropymain} achieves its maximum value, $ \frac{1}{2}\sqrt{d_S}$ when the effective environment's dimension is one, $\dim{\mc{H}_N}=1$. In this case, no information about the microscopic description is discarded; the subsystem is the full system. The most common example is that of a unitary channel, as in a closed dynamics, and thus with zero entropy. Considering the general $d_R$-depolarizing channel $\Lambda_{d_R}^\lambda(\psi)=\lambda \frac{\mathds{1}_R}{d_R} + (1-\lambda)\psi$, with $0 \leq \lambda \leq 1 + 1/(d_R^2-1)^2$, the bound in \eqref{eq:ineqentropymain} is $\frac{1}{2}\sqrt{\frac{\lambda(2-\lambda)}{d_R}+d_R(1-\lambda)^2}$ (proof in Section V of Supplemental Material). As $\lambda$ increases to 1, the channel approaches the maximum entropy channel, thus leading to the smallest value for the bound, $\frac{1}{2}\sqrt{1/d_R}$. Equation \eqref{eq:ineqentropymain} reduces to Eq. \eqref{eq:popescubound} in the special case where $\Lambda=\tr_E$ (proof in Section III of Supplemental Material).

Broadly speaking,  the average distance between a state $\rho_\Lambda^\psi=\Lambda(\proj{\psi})$, for $\ket{\psi}$ sampled uniformly from $\mathbb{S}^{2d_R-1}$, to the canonical state $\Omega_\Lambda=\Lambda(\idty_R/d_R)$ will be smaller the more information about $\proj{\psi}$ is thrown away by $\Lambda$, i.e., the higher is the channel entropy. 

Another way to understand the result in Eq.~\eqref{eq:ineqentropymain}, is by explicitly using the expression for $\Lambda$, Eq.~\eqref{eq:channel}:
\[
S_L(\Lambda)= 1 - \tr\left\{ \left[ \tr_N( V\otimes\idty \proj{\phi_+}V^\dagger\otimes \idty  )\right]^2 \right\};
\]
which can be interpreted in terms of the entanglement in  $V\otimes\idty \ket{\phi_+}\in (\mc{H}_S\otimes\mc{H}_N)\otimes\mc{H}_R$ with respect  to the partition that splits the effective environment against the rest. If the effective environment  is separable from the effective system and the copy in $\mc{H}_R$,  then no information is lost by tracing it out. In the other direction, the bound in \eqref{eq:ineqentropymain} is tighter the stronger is the entanglement, generated by the isometry $V$, between the effective system and its effective environment.

The second part of the canonical typicality result follows from the application of Levy's lemma to the distance between $\Lambda(\proj{\psi})$ and $\Omega_\Lambda$. For  any state uniformly sampled from $\mathbb{S}^{2d_\text{R} -1}$ it follows that:
\begin{equation}
\label{eq:levycanonical}
\pr_{\psi}\Big[\big|\mathcal{D}\left(\Lambda(\proj{\psi}),\Omega_\Lambda\right)- \overline{\mathcal{D}(\rho_\Lambda^\chi,\Omega_\Lambda)}^\chi\big|>\epsilon \Big] \le 2 \textrm{e}^{-C\frac{ d_\text{R}\epsilon^2}{4\eta_\Lambda^2}}.
\end{equation}
In the above expression $C$ is a constant that can be taken equal to $2/9\pi^3$, and $\eta_\Lambda:= \max_{\rho,\sigma} \norm{\Lambda(\rho)-\Lambda(\sigma)}_1/\norm{\rho-\sigma}_1$ is the channel Lipschitz constant. See Section I of Supplemental Material for details.

From the first part of the result, Eq.~\eqref{eq:ineqentropymain}, we have that if the channel's entropy is large, the average distance between $\Lambda(\proj{\psi})$ and $\Omega_\Lambda$ is small. Putting this together with Eq.~\eqref{eq:levycanonical}, the probability for  $\mathcal{D}\left(\Lambda(\proj{\psi}),\Omega_\Lambda\right)$ be further from its mean value by an amount $\epsilon$ decays exponentially with $\epsilon^2$ and $d_R$. In this generalized scenario, the conditions for the canonical typicality are then $S_L(\Lambda)\approx 1$ and $d_R\gg 1$.

\textit{Application: Blurred and saturated detector.} Consider hundreds of cold atoms loaded in an optical lattice. Nowadays, this system can be highly isolated from external environments during the experiment. However, fully describing its quantum state may be experimentally unfeasible, and theoretically intractable. It is thus highly desirable to obtain effective descriptions which are more manageable while not losing experimental relevance.

Concretely, consider $N$ two-level atoms in an optical lattice, one atom per potential well, and no interactions. The Hamiltonian describing this situation is then $H=-\hbar\omega\sum_{i=1}^{N} Z_i/2$, with $Z_i$ the usual $z$ Pauli matrix for the $i$-th atom. Suppose that due to a energy restriction, a fraction $p$ of the atoms are excited, $\ket{1}$, and the rest are in the ground state, $\ket{0}$. Accordingly, $\mc{H}_R\subseteq \mc{H}_T=(\Cx_2)^{\otimes N}$ has dimension $d_R={N \choose Np}$. The microcanonical state for this scenario is simply $\mc{E}_R=\sum_{s:|s|=Np}\proj{s}/d_R$. Here, $s\in\{0,1\}^N$ are strings with $N$ bits, and $|s|$ represents the number of excited atoms, i.e. the number of 1's in $s$.

In this type of experiment, the energy measurement of the atoms is frequently done with  a fluorescence technique: over the system is shone a laser whose frequency is chosen to be resonant with a transition of the excited state with a third level (this level is only used in the measurement process, but not to encode information). This level is broad and the electron quickly decays back to $\ket{1}$ by emitting a photon in a random direction, and the process repeats. In this way, if a given atom is in  $\ket{1}$, it will scatter light. On the contrary, if the atom is in $\ket{0}$, the laser is far from resonance and no light is scattered. The light scattered by the various atoms is collected by a microscope whose resolution determines if the light scattered by neighboring atoms can be resolved.

Consider a situation where the fluorescence measurement cannot resolve individual wells, but it takes blocks of $n$ sites. In this way, $n$ atoms behave as a single effective atom: if at least one atom is excited, light will be scattered; if no atom is excited, no light is scattered. Which canonical state should be assigned to this system? This situation is not the usual system-environment split, but our formalism can be employed to construct the generalized canonical state.  

To describe this scenario, inspired by previous works~\cite{cris2017, oleg,pedrinho, carlospineda2021}, we define the CPTP map $\Lambda_\text{BnS}^{n\rightarrow 1}:\mc{L}(\Cx_2^{\otimes n})\mapsto\mc{L}(\Cx_2)$: 
\[
\Lambda_\text{BnS}^{n\rightarrow 1}[\ket{s'}\!\bra{s''}]=\left\{\begin{aligned}
	&\proj{0}, \hspace*{0.2cm} \text{if } s'=s'' \text{ and } |s'|= 0,\\
	&\proj{1}, \hspace*{0.2cm} \text{if } s'=s'' \text{ and } |s'|\ge 1,\\
	&\frac{1}{\sqrt{2^n-1}}\ket{0}\!\bra{1}, \hspace*{0.2cm} \text{if } |s'|=0 \text{ and } |s''|\ge 1,\\
	&\frac{1}{\sqrt{2^n-1}}\ket{1}\!\bra{0}, \hspace*{0.2cm} \text{if } |s'|\ge 1 \text{ and } |s''| = 0,\\
	& 0, \hspace*{0.2cm} \text{otherwise. }
\end{aligned}
\right.
\]
In the above, $s', s''\in\{0,1\}^n$ are  strings with $n$ bits.  The factor of $1/\sqrt{2^n-1}$ in the coherence terms is the largest possible while keeping $\Lambda_\text{BnS}^{n\rightarrow 1}$ completely positive. There is no effective coherence  from terms as $\ket{s'}\!\bra{s''}$ with $s'\neq s''$ and $|s'|,|s''|\ge 1$, because these states cannot be distinguished by the detection process. Notice that such a map is not equivalent to neither a partial trace nor to an unitary reshuffle followed by partial trace, as the result $\proj{1}$ is obtained if at least one atom is excited independently of its position~\footnote{This can be easily apprehended in the simple case where two atoms are seeing as a single effective atom due to the lack of resolution in the measurement process, $\Lambda_\text{BnS}^{2\rightarrow 1}$. In this case we have the following map: $\proj{01} \mapsto \proj{1}$, $\proj{10} \mapsto \proj{1}$, and $\proj{11} \mapsto \proj{1}$. Clearly this cannot be cast as a unitary transformation followed by the partial trace of either one of the particles.}. 

Therefore, if we split the $N$  atoms of the total system into $k=N/n$ blocks of $n$ atoms, we can obtain a generalized effective subsystem description via the map:
\begin{equation}
	\Lambda_\text{BnS}^{N\rightarrow k}=\Lambda_\text{BnS}^{\frac{N}{k}\rightarrow 1}\otimes\Lambda_\text{BnS}^{\frac{N}{k}\rightarrow 1}\otimes\cdots\otimes\Lambda_\text{BnS}^{\frac{N}{k}\rightarrow 1}. \nonumber
\end{equation}
The effective system is then equivalent to  $k$  two-level atoms.

For this scenario, we can evaluate the generalized canonical state, $\Omega_{\Lambda_\text{BnS}}=\Lambda_\text{BnS}^{N\rightarrow k}(\mc{E}_R)$, to be:
\begin{equation}
\Omega_{\Lambda_\text{BnS}}=\frac{1}{d_R}\sum_{|s|=\ceil{k p}}^{\min(Np, k)} \sum_{q=0}^{|s|} {|s| \choose q}{ \frac{N}{k}q \choose Np} (-1)^{|s|-q}\; \Pi_{|s|}.\nonumber
\end{equation}
In the above expression $\Pi_{|s|}$ is the projector onto the subspace spanned by the strings with number of 1's equal $|s|$. The derivation is shown in Section IV of Supplemental Material.

Given our results, we expect to observe the canonical typicality whenever  $d_R\gg d_S=2^k$. Within these assumptions, $\Lambda_\text{BnS}^{N\rightarrow k}$ will disregard a considerable amount of information about the microscopic description, i.e., $S(\Lambda_\text{BnS}^{N\rightarrow k})\approx 1$. In this case, selecting any state at random in $\mc{H}_R$ and applying $\Lambda_\text{BnS}^{N\rightarrow k}$ to it will lead to a good approximation of $\Omega_{\Lambda_\text{BnS}}$.

\begin{figure}[h!]
	\centering
	\begin{tabular}{l}
		(a)\\
		\includegraphics[width=8.5cm,valign=t]{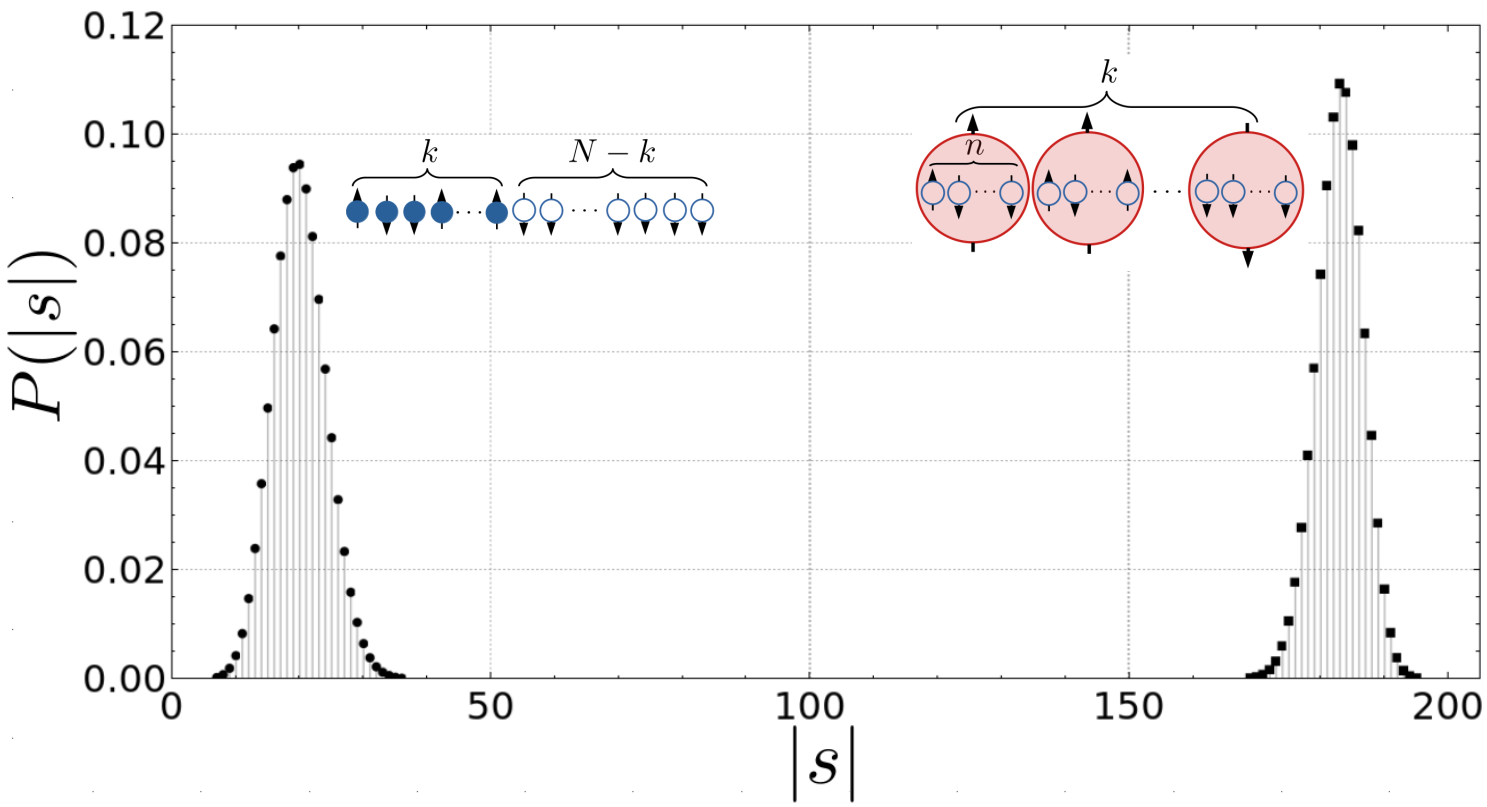}\\
		(b)\\
		\includegraphics[width=8.4cm,valign=t]{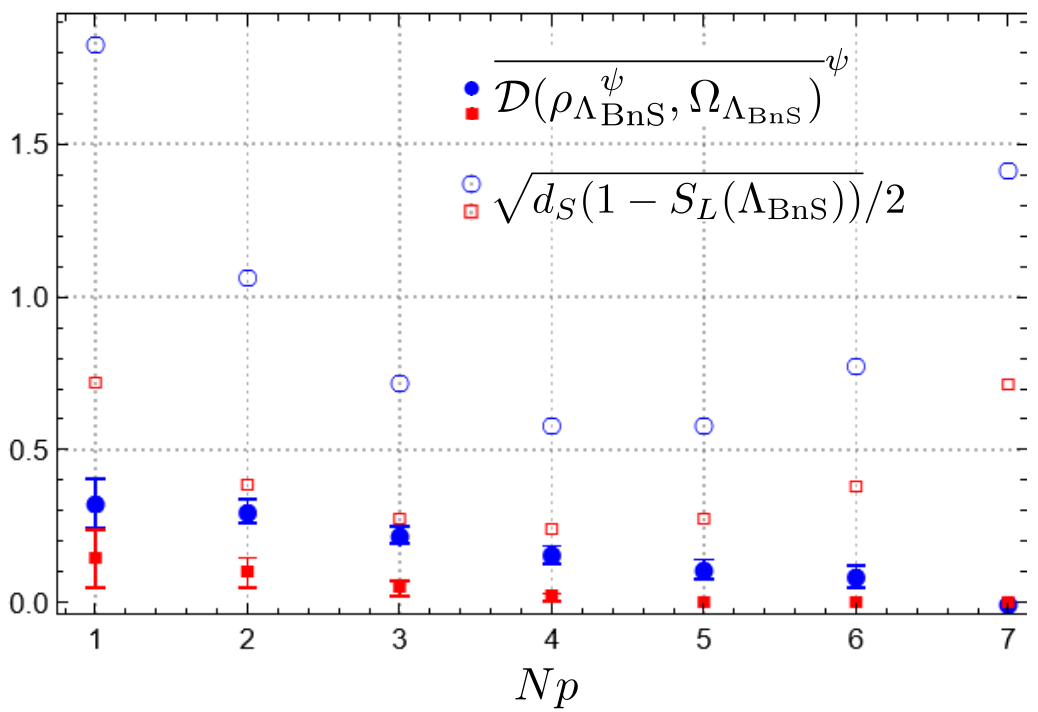}

	\end{tabular}
	\caption{ \textbf{(a) Canonical Energy Distributions.} Probability distribution of $|s|$ (which is proportional to the system energy) for two scenarios: the points on the left (circles) represent the energy distribution for the partial trace case, while the points on the right (squares) are related to the blurred and saturated situation. For this plot we used $ N=10000 $, $ Np=200 $ and $k=1000$. \textbf{(b) Canonical typicality for the blurred and saturated generalized subsystem.} Filled markers represent the average distance to the generalized canonical state, and the hollow markers show the bound in Eq.~\eqref{eq:ineqentropymain}. The underlying system is composed by $N=8$ spins, with the red (squares) points representing the case where $k=2$ and the blue (circles) points the case $k=4$.
	}  
	\label{fig:BnSStatistics}
\end{figure}

Fixed the specified Hamiltonian, in Fig.\ref{fig:BnSStatistics} we compare two scenarios with a subsystem of $k$ spins: the first one we simply traced out $N-k$ spins, i.e., we apply $\Lambda_{\tr_{N-k}}$. The second one is the scheme described above, with the subsystem of $k$ spins being obtained by applying the map $\Lambda_{\textrm{BnS}}^{N\rightarrow k}$. 

First, note that the canonical states for the two situations are very different. Noticing that the energy of the system is proportional to $|s|$, in Fig.~\ref{fig:BnSStatistics}(a) we plot the energy distribution assuming $N=10000$, $N p = 200$ and $k=100$ for both cases. The distribution  more to the left is the one associated with the partial trace, whereas the one more to the right is related to the blurred and saturated detection. From this result it is clear that the appropriate canonical state heavily depends on the generalized subsystem description, and not exclusively on the underlying particle structure.

The second point concerns the canonical typicality for the blurred and saturated detection, as this aspect for the  partial trace case has been already explored in~\cite{popescu2006}. In Fig.~\ref{fig:BnSStatistics}(b) the solid symbols refer to the mean and variance for the distribution of $\mathcal{D}(\rho_{\Lambda_\textrm{BnS}}^\psi,\Omega_{\Lambda_\textrm{BnS}})$ with $\psi$ taken at random from the uniform measure, while the hollow markers show the bound $\sqrt{d_S(1-S_L(\Lambda_\textrm{BnS}))}/2$. In this plot, the underlying system is assumed to be formed by 8 spins, $N=8$, and the case where $k=4$ is shown in blue (circles), while the case of $k=2$ is shown in red (squares). All the quantities are shown as a function of the number of excitations, $Np$, which is directly linked to the restriction dimension $d_R$. One immediately observes  that as $k$ decreases, the smaller are the mean distances and the bound. Also it is clear that the bound is tighter the bigger is $d_R$. Both behaviors are can be understood by noticing that, for fixed $N$, the smaller $k$  and the larger $d_R$ are, the more information about the microscopic state is being thrown away, i.e., the larger is the channel entropy. Lastly, from the error bars,  the concentration around the mean value gets stronger as the fraction of excited spins increases. In Eq.~\eqref{eq:levycanonical} this is expressed by the Lipschitz constant of $\Lambda_{\text{BnS}}$, which is smaller than one and decreases as  $Np$ grows. For instance, for $Np=7$ no block of  4 (or 2) spins will have all the spins in the ground state, and thus the only possible state for the $k$ effective spins is to have all of them in the excited state. Therefore, all pure states in the subspace with $Np=7$ are mapped to $k$ excited effective spins and the canonical state is also given by $k$ excited state, and as such their distance and $\eta_{\Lambda_{\textrm{BnS}}}$ are both zero. 

\textit{Conclusions.} This work is a first step toward a statistical mechanics of generalized subsystems. We introduced the canonical state for generalized subsystems and showed that there are situations where it is drastically different from the canonical state one would obtain by associating the relevant degrees of freedom of a physical scenario with the system's particles. We also showed that the phenomena of canonical typicality is also present for generalized subsystems, and thus key ingredients of traditional statistical mechanics are inherited by this new picture. Given our all-encompassing description of physical systems, it became explicit that the quantity which controls the emergence of canonical typicality is the entropy of the channel used to define the generalized subsystem.

One last remark: in the presented formalism we employed quantum channels exclusively to define the generalized subsystems. However, given the formalism's flexibility, we could  change $\Lambda$ by a composition of channels as $\Lambda\circ\Gamma$, with $\Gamma:\mc{L}(\mc{H}_R) \mapsto \mc{L}(\mc{H}_R)$, without altering any of the presented results. In this situation $\Gamma$ can be thought as a pre-processing of the underlying description, with  pure states of $\mc{H}_R$ generically being  mapped into mixed ones. Such a possibility is interesting in at least two ways. First it shows that the phenomena of canonical typicality is robust to noise, which might substantiate  the success of ensembles. Second, this might alleviate the requirement for canonical typicality that the underlying states whose effective description behave similarly to the canonical state, are states taken from the uniform measure. For complex systems we do not expect nature to be able to efficiently sample from such a measure~\cite{verstraete}. By preprocessing the states from the uniform measure, we are effectively inducing a new measure, which might lead to more physically feasible samples. For instance, it is expected that the average amount of entanglement to drop after the pre-processing.

Altogether, the presented formalism not only gives a more suitable description of the statistical mechanics of complex quantum systems, but it also sheds light on some conceptual issues of traditional statistical mechanics.

\textit{Acknowledgments.} This work is supported in part by the National Council for Scientific and Technological Development, CNPq Brazil (projects: Universal Grants No. 406499/2021-7, and No. 409611/2022-0), and it is part of the Brazilian National Institute for Quantum Information. T.R.O. acknowledges funding from the Air Force Office of Scientific Research under Grant No. FA9550-23-1-0092.

%\bibliographystyle{apsrev4-1}	
%\bibliography{ref} 

%

\onecolumngrid
\appendix
\section{Supplemental Material}
\section{I. Applying Levy's Lemma to $\mathcal{D}(\rho_\Lambda^\psi,\Omega_\Lambda)$}

\textbf{Lemma \textit{(Levy's Lemma)}:} \textit{Given a Lipschitz continuous function $f: \mathbb{S}^d\rightarrow\mathbb{R}$, and a point $\psi\in\mathbb{S}^d$ chosen uniformly at random,}
\begin{align}
	\pr_\psi\big[ | f(\psi) - \overline{ f }|  \geq \epsilon \big] \leq 2\exp\bigg(\dfrac{-2G(d+1)\epsilon^2}{\eta^2}\bigg),
	\label{eq:lemma}
\end{align}
\textit{where $\eta$ is the Lipschitz constant of $f$ with respect to the Euclidean norm (see below),  $G$ is a positive constant (which  can be taken to be $G=(18\pi^3)^{-1}$), and $\overline{f}$ is the mean value of $f$ over the uniform measure in $\mathbb{S}^d$.}

After noticing that the space of pure quantum states in a $d$-dimensional Hilbert space ($\mc{H}_d$) can be described as a  hyper-sphere embedded in a $2d$ real space, $\mathbb{S}^{2d-1}$, we can apply Levy's Lemma to Lipschitz continuous properties of pure states by sampling them from the Haar measure.

In our case, we are interested in applying Levy's Lemma to $f(\psi)= \mathcal{D}(\rho_\Lambda^\psi,\Omega_\Lambda)$ with $\psi \in \mathbb{S}^{2d_R-1}$. It remains to show that such a function is Lipschitz continuous with respect to the Euclidean norm.

Generally, a function $f:X\mapsto Y$ between metric spaces is Lipschitz continuous if  there exists a constant $\eta\in\Rl^+$ such that for all $a,b\in X$ we have  $\norm{f(a)-f(b)}_Y\le \eta \norm{a-b}_X$. Here $\norm{\cdot}_Z$ is a  norm associated with space $Z$. Lispschitz continuity is then a strong form of continuity.

For our case, we must upper bound $|\mathcal{D}(\rho_\Lambda^\psi,\Omega_\Lambda)- \mathcal{D}(\rho_\Lambda^\phi,\Omega_\Lambda)|$ for any $\psi$ and $\phi$ in $\mathbb{S}^{2d_R-1}$. It goes as follows [25]:
\begin{align}
	|\mathcal{D}(\rho_\Lambda^\psi,\Omega_\Lambda)- \mathcal{D}(\rho_\Lambda^\phi,\Omega_\Lambda)|^2&= | \norm{\Lambda(\psi)-\Lambda(\mc{E}_R)}_1 -\norm{\Lambda(\phi)-\Lambda(\mc{E}_R)}_1|^2/4;\nonumber\\ 
	&\le  \norm{\Lambda(\psi)-\Lambda(\phi)}_1^2/4;\nonumber\\
	&\le \eta_\Lambda^2 \norm{\proj{\psi}-\proj{\phi}}_1^2/4;\nonumber\\
	&= \eta_\Lambda^2 (1- |\<\psi|\phi\>|^2);\nonumber\\
	&\le \eta_\Lambda^2\norm{\ket{\psi}-\ket{\phi}}^2.
	\label{eq:lip}
\end{align}
This proves that $f(\psi)= \mathcal{D}(\rho_\Lambda^\psi,\Omega_\Lambda)$ is Lipschitz continuous with constant $\eta_\Lambda:= \max_{\rho,\sigma} \norm{\Lambda(\rho)-\Lambda(\sigma)}_1/\norm{\rho-\sigma}_1$.

We then obtain the desired result:
\begin{equation}
	\pr_{\psi}\Big[\big|\mathcal{D}\left(\Lambda(\proj{\psi}),\Omega_\Lambda\right)- \overline{\mathcal{D}(\rho_\Lambda^\chi,\Omega_\Lambda)}^\chi\big|\ge\epsilon \Big] \le 2 \textrm{e}^{-C\frac{ d_\text{R}\epsilon^2}{\eta_\Lambda^2}}.
\end{equation}
where $C$ is a constant that can be taken equal to $2/9\pi^3$, and $\eta_\Lambda$ is the channel Lipschitz constant. It is worthy mentioning that $\eta_\Lambda$ may depend on the channel properties, and it can be evaluated for each choice of channel. Nevertheless, from its definition,  it cannot be bigger than 1, which can then be used as a channel-independent value.

\section{II. Proving $ \overline{\mathcal{D}(\rho_\Lambda^\psi,\Omega_\Lambda)}^\psi\leq\frac{1}{2} \sqrt{d_S\tr({J_\Lambda}^2)}=\frac{1}{2} \sqrt{d_S(1-S_L(\Lambda))}$}
\label{app:mean_dist}

Given that the definitions of the trace norm $\parallel A \parallel_1=\tr\big[\sqrt{A^\dagger A}\big]$ and Hilbert-Schmidt norm $\parallel A\parallel_2=\sqrt{\tr[A^\dagger A]}$. For any $n\times n$ matrix $A$ the following inequality is satisfied:
\begin{align}
	\parallel A \parallel_1\leq\sqrt{n}\parallel A \parallel_2.
	\label{eq:nomsrelation}
\end{align}
This relation can be easily proved, considering that $A$ has eigenvalues $\lambda_i$, and the convexity of the square function:
\begin{align}
	\parallel A \parallel_1^2=n^2\bigg(\dfrac{1}{n}\sum_i|\lambda_i|\bigg)^2 \nonumber \leq n^2\dfrac{1}{n}\sum_i|\lambda_i|^2=n\parallel A \parallel_2^2.
\end{align}

Let us derive another useful relation. Doing $ A=\rho_\Lambda^\psi-\Omega_\Lambda $ and taking the average in the Hilbert-Schmidt norm: 
\begin{align}
	\overline{\parallel \rho_\Lambda^\psi-\Omega_\Lambda \parallel_2}^\psi &\leq\sqrt{\overline{ \parallel \rho_\Lambda^\psi-\Omega_\Lambda \parallel_2^2}^\psi} \nonumber \\
	&=\sqrt{\overline{\tr[(\rho_\Lambda^\psi-\Omega_\Lambda)^2]}^\psi} = \sqrt{\overline{\tr[(\rho_\Lambda^\psi)^2]}^\psi-2\overline{\tr[\rho_\Lambda^\psi\Omega_\Lambda]}^\psi+\overline{\tr[\Omega_\Lambda^2]}^\psi} = \sqrt{\overline{\tr[(\rho_\Lambda^\psi)^2]}^\psi-2\tr[\overline{\rho_\Lambda^\psi}^\psi\Omega_\Lambda]+\tr[\Omega_\Lambda^2]} \nonumber \\
	&=\sqrt{\overline{\tr[(\rho_\Lambda^\psi)^2]}^\psi-\tr[\Omega_\Lambda^2]},
	\label{eq:relation2}
\end{align}
in the first line we used the Jensen's inequality $f( \overline{M })\leq\overline{ f(M) }$ with $ f(\cdot)=(\cdot)^2 $. In the third line we used $ \overline{\Omega_\Lambda}^\psi=\overline{\rho_\Lambda^\psi}^\psi=\Omega_\Lambda $. 

Then, from (\ref{eq:nomsrelation}) and (\ref{eq:relation2}):
\begin{align}
	\overline{\mathcal{D}(\rho_\Lambda^\psi,\Omega_\Lambda)}^\psi=\dfrac{1}{2}\overline{\parallel \rho_\Lambda^\psi-\Omega_\Lambda \parallel_1}^\psi\leq\dfrac{1}{2}\sqrt{d_S\big(\overline{\tr[(\rho_\Lambda^\psi)^2]}^\psi-\tr[\Omega_\Lambda^2]\big)}.
	\label{eq:ineq1}
\end{align}

To calculate $\overline{\tr[(\rho_\Lambda^\psi)^2]}^\psi$, it is helpful to duplicate the original Hilbert space, extending the problem from $\mathcal{H}_R$ to $\mathcal{H}_R\otimes\mathcal{H}_{R'}$. We start by written $\tr[(\rho_\Lambda^\psi)^2]$ as:
\begin{align}
	\tr[(\rho_\Lambda^\psi)^2]&=\sum_{i}(\rho_{ii})^2 =\sum_{i,j,i',j'}(\rho_{ij})(\rho_{i'j'})\langle ii'|jj'\rangle\langle i'i|jj'\rangle \nonumber \\
	&=\tr_{SS'}[(\rho_\Lambda^\psi\otimes\rho_{\Lambda'}^\psi)F_{SS'}], 	
\end{align}
where the $F_{SS'}$ is the swap operation $S\leftrightarrow S'$:
\begin{align}
	F_{SS'}=\sum_{ss'}\ket{s'}\bra{s}\otimes\ket{s}\bra{s'}
	\label{eq:swap}
\end{align}
and then
\begin{align}
	\overline{\tr[(\rho_\Lambda^\psi)^2]}^\psi&=\overline{\tr_{SS'}[(\rho_\Lambda^\psi\otimes\rho_{\Lambda'}^\psi)F_{SS'}]}^\psi \nonumber \\
	&=\overline{\tr_{SS'}[(\Lambda[\ket{\psi}\bra{\psi}]\otimes\Lambda[\ket{\psi}\bra{\psi}])F_{SS'}]}^\psi \nonumber \\
	&=\overline{\tr_{SS'}[(\Lambda\otimes\Lambda)[\ket{\psi}\bra{\psi}\otimes\ket{\psi}\bra{\psi}]F_{SS'}]}^\psi \nonumber \\
	&=\tr_{SS'}[(\Lambda\otimes\Lambda)[\overline{\ket{\psi}\bra{\psi}\otimes\ket{\psi}\bra{\psi}}^\psi]F_{SS'}]. 
	\label{eq:rho2}
\end{align}
Using the result [8]
\begin{align}
	\overline{\ket{\psi}\bra{\psi}\otimes\ket{\psi}\bra{\psi}}^\psi=\dfrac{\mathds{1}_{RR'}+F_{RR'}}{d_R(d_R+1)}
\end{align}
where $\mathds{1}_{RR'}=\mathds{1}_R\otimes\mathds{1}_{R'}$. Using above equation in (\ref{eq:rho2}):
\begin{align}
	\overline{\tr[(\rho_\Lambda^\psi)^2]}^\psi&=\tr_{SS'}\big[(\Lambda\otimes\Lambda)\big[\dfrac{\mathds{1}_{RR'}}{d_R(d_R+1)}\big]F_{SS'}\big] + \tr_{SS'}\big[(\Lambda\otimes\Lambda)\big[\dfrac{\mathds{F}_{RR'}}{d_R(d_R+1)}\big]F_{SS'}\big] \nonumber \\
	&\leq\tr_S[\Omega_\Lambda^2]+\tr_{SS'}\big[(\Lambda\otimes\Lambda)\big[\dfrac{F_{RR'}}{d_R^2}\big]F_{SS'}\big]
	\label{eq:rho22}
\end{align}
in the last line we used $d_R(d_R+1)\geq d_R^2$.

Thus, (\ref{eq:rho22}) in (\ref{eq:ineq1}):
\begin{align}
	\overline{\parallel \rho_\Lambda^\psi-\Omega_\Lambda \parallel_1}^\psi\leq\sqrt{\dfrac{d_S}{d_R^2}\;\tr_{SS'}[(\Lambda\otimes\Lambda)[F_{RR'}]F_{SS'}]}.
	\label{eq:ineq2}
\end{align}
Using (\ref{eq:swap}) in $\tr_{SS'}[(\Lambda\otimes\Lambda)[F_{RR'}]F_{SS'}]$:
\begin{align}
	\tr_{SS'}[(\Lambda\otimes\Lambda)[F_{RR'}]F_{SS'}]&=\tr_{SS'}\big[(\Lambda\otimes\Lambda)\big[\sum_{r,r'=0}^{d_R-1}\ket{r'}\bra{r}\otimes\ket{r}\bra{r'}\big] F_{SS'}\big] \nonumber \\
	&=\tr_{SS'}\big[\big(\sum_{r,r'=0}^{d_R-1}\Lambda\big[\ket{r'}\bra{r}\big]\otimes\Lambda\big[\ket{r}\bra{r'}\big]\big) F_{SS'}\big].
	\label{eq:intermed1} 
\end{align}

Any quantum channel $\Lambda:\mathcal{L}(\mathcal{H}_R)\rightarrow\mathcal{L}(\mathcal{H}_S)$, there exist a representation
\begin{align}
	\Lambda[\psi]=\sum_{m=1}^{\tau} K_m\psi K_m^\dagger
	\label{eq:krausrep}
\end{align} 
for all $\psi \in \mathcal{L}(\mathcal{H}_R)$, $K_m\in\mathcal{L}(\mathcal{H}_R,\mathcal{H}_S)$ for all $m\in\{1,\dots,\tau\}$ and $\tau$ need not be any larger than $d_Rd_S$. The expression (\ref{eq:krausrep}) is a \textit{Kraus representation} of the map $\Lambda$, unlike the Choi representation, Kraus representations are not unique.

Using (\ref{eq:swap}) and (\ref{eq:krausrep}) in (\ref{eq:intermed1})

\begin{align}
	\tr_{SS'}[(\Lambda\otimes\Lambda)[F_{RR'}]F_{SS'}]&=\tr_{SS'}\big[\big(\sum_{r,r'=0}^{d_R-1}\sum_{m,n=1}^{\tau}K_m\ket{r'}\bra{r}K_m^\dagger\otimes K_n\ket{r}\bra{r'}K_n^\dagger\big)\sum_{s,s'=0}^{d_S-1} \ket{s'}\bra{s}\otimes\ket{s}\bra{s'}\big]  \nonumber \\
	&=\tr_{SS'}\big[\sum_{r,r'=0}^{d_R-1}\sum_{m,n=1}^{\tau}\sum_{s,s'=0}^{d_S-1}K_m\ket{r'}\bra{r}K_m^\dagger\ket{s'}\bra{s}\otimes K_n\ket{r}\bra{r'}K_n^\dagger\ket{s}\bra{s'}\big]  \nonumber \\
	&=\sum_{t,t'=0}^{d_S-1}\sum_{r,r'=0}^{d_R-1}\sum_{m,n=1}^{\tau}\sum_{s,s'=0}^{d_S-1}\bra{t}K_m\ket{r'}\bra{r}K_m^\dagger\ket{s'}\overbrace{\bra{s}t}^{\delta_{st}}\rangle\bra{t'} K_n\mathds{1}_{R'}\ket{r}\bra{r'}K_n^\dagger\ket{s}\overbrace{\bra{s'}t'}^{\delta_{s't'}}\rangle \nonumber \\
	&=\sum_{r,r'=0}^{d_R-1}\sum_{m,n=1}^{\tau}\sum_{s,s'=0}^{d_S-1}\bra{s}K_m\ket{r'}\bra{r}K_m^\dagger\ket{s'}\bra{s'} K_n\ket{r}\bra{r'}K_n^\dagger\ket{s} \nonumber \\
	&=\sum_{m,n=1}^{\tau}\sum_{s,s'=0}^{d_S-1}\bra{s}K_m\underbrace{\big(\sum_{r'=0}^{d_R-1}\ket{r'}\bra{r'}\big)}_{=\mathds{1}_R}K_n^\dagger\ket{s}\bra{s'} K_n\underbrace{\big(\sum_{r=0}^{d_R-1}\ket{r}\bra{r}\big)}_{=\mathds{1}_R}K_m^\dagger\ket{s'} \nonumber \\
	&=\sum_{m,n=1}^{\tau}\sum_{s,s'=0}^{d_S-1}\bra{s}K_mK_n^\dagger\ket{s}\bra{s'} K_nK_m^\dagger\ket{s'} \nonumber \\
	&=\sum_{m,n=1}^{\tau}\tr[K_mK_n^\dagger]\tr[ K_nK_m^\dagger]=\sum_{m,n=1}^{\tau}|\tr[K_mK_n^\dagger]|^2,
\end{align}
above result in (\ref{eq:ineq2}), we have:
\begin{align}
	\overline{\mathcal{D}(\rho_\Lambda^\psi,\Omega_\Lambda)}^\psi\leq\dfrac{1}{2}\sqrt{\dfrac{d_S}{d_R^2}\sum_{m,n=i}^{\tau}\big|\tr\big[K_mK_n^\dagger\big]\big|^2}.
	\label{eq:ineqkraus}
\end{align}

Rewriting the above equation in terms of Choi-Jamiołkowski state, $J_\Lambda:=\Lambda\otimes\idty(\proj{\phi^+})$ of the quantum channel $\Lambda$, where $\ket{\phi^+}=\sum_i\ket{ii}/\sqrt{d_R}\in\mc{H}_R\otimes\mc{H}_{R'}$. Using the Krauss form:
\begin{align}
	J_\Lambda&=(\Lambda\otimes\mathds{1}_R)[\ket{\phi_+}\bra{\phi_+}] \nonumber \\
	&=\dfrac{1}{d_R}\sum_{i,j=0}^{d_R-1}\Lambda[\ket{i}\bra{j}]\otimes\ket{i}\bra{j}=\dfrac{1}{d_R}\sum_{i,j=0}^{d_R-1}\sum_{m=1}^{\tau}K_m\ket{i}\bra{j}K_m^\dagger\otimes\ket{i}\bra{j}.
\end{align}

Calculating ${d_R}^2\tr[{J_\Lambda}^2]$:
\begin{align}
	{d_R}^2\tr[{J_\Lambda}^2]&=\tr_{SR}\bigg[\big(\sum_{i,j=0}^{d_R-1}\sum_{m=1}^{\tau}K_m\ket{i}\bra{j}K_m^\dagger\otimes\ket{i}\bra{j}\big)\big(\sum_{i',j'=0}^{d_R-1}\sum_{n=1}^{\tau}K_n\ket{i'}\bra{j'}K_n^\dagger\otimes\ket{i'}\bra{j'}\big)\bigg] \nonumber \\
	&=\tr_{SR}\bigg[\sum_{i,j,i',j'}\sum_{m,n}K_m\ket{i}\bra{j}K_m^\dagger K_{n}\ket{i'}\bra{j'}K_{n}^\dagger\otimes\ket{i}\overbrace{\bra{j}i'\rangle}^{\delta_{ji'}}\bra{j'}\bigg] \nonumber \\
	&=\tr_{SR}\bigg[\sum_{i,j,j'}\sum_{m,n}K_m\ket{i}\bra{j}K_m^\dagger K_{n}\ket{j}\bra{j'}K_{n}^\dagger\otimes\ket{i}\bra{j'}\bigg] \nonumber \\
	&=\sum_{s=0}^{d_S-1}\sum_{i,j,j'}\sum_{m,n}\bra{s}K_m\ket{i}\bra{j}K_m^\dagger K_{n}\ket{j}\bra{j'}K_{n}^\dagger\ket{s}\overbrace{\sum_{r=0}^{d_R-1}\langle r\ket{i}\bra{j'}r\rangle}^{\delta_{j'i}} \nonumber \\
	&=\sum_{s}\sum_{i,j}\sum_{m,n}\bra{s}K_m\ket{i}\bra{j}K_m^\dagger K_{n}\ket{j}\bra{i}K_{n}^\dagger\ket{s} \nonumber \\
	&=\sum_{s}\sum_{j}\sum_{m,n}\bra{s}K_m\bigg(\sum_{i}\ket{i}\bra{i}\bigg)K_{n}^\dagger\ket{s}\bra{j}K_m^\dagger K_{n}\ket{j} \nonumber \\
	&=\sum_{s}\sum_{j}\sum_{m,n}\bra{s}K_mK_{n}^\dagger\ket{s}\bra{j}K_m^\dagger K_{n}\ket{j} \nonumber \\
	&=\sum_{m,n}\bigg(\sum_{s}\bra{s}K_mK_{n}^\dagger\ket{s}\bigg)\bigg(\sum_{j}\bra{j}K_m^\dagger K_{n}\ket{j}\bigg) \nonumber \\
	&=\sum_{m,n}\tr\big[K_mK_{n}^\dagger\big]\overbrace{\tr\big[K_m^\dagger K_{n}\big]}^{\tr\big[K_nK_{m}^\dagger\big]}=\sum_{m,n=1}^{\tau}|\tr[K_mK_n^\dagger]|^2.
\end{align}
So (\ref{eq:ineqkraus}) can be written in terms of the purity of the Choi-Jamiołkowski state
\begin{align}
	\overline{\mathcal{D}(\rho_\Lambda^\psi,\Omega_\Lambda)}^\psi\leq\dfrac{1}{2}\sqrt{d_S\tr\big[{J_\Lambda}^2\big]},
	\label{eq:ineqchoi}
\end{align}

or in terms of the linear entropy of the map: 
\begin{align}
	\overline{\mathcal{D}(\rho_\Lambda^\psi,\Omega_\Lambda)}^\psi\leq\dfrac{1}{2}\sqrt{d_S(1-S_L(\Lambda))},
	\label{eq:ineqentropy}
\end{align}
since $ S_L(\Lambda) = 1 - \tr\big[{J_\Lambda}^2\big] $.

\section{III. Demonstrating that $\sqrt{d_S\tr(J_{\tr_E}^2)}=\sqrt{\frac{d_S}{d_E^{eff}}}$}

This section aims to demonstrate that $\overline{\mc{D}(\rho_{\tr_E}^\psi,\Omega_{\tr_E})}^\psi \le \frac{1}{2}\sqrt{d_S/d_E^\text{eff}}$ is a particular case of $\overline{\mathcal{D}(\rho_\Lambda^\psi,\Omega_\Lambda)}^\psi\leq\dfrac{1}{2}\sqrt{d_S\tr\big[{J_\Lambda}^2\big]}$, when $ \Lambda = \tr_E $. So we need to show that $ \tr(J_{\tr_E}^2) = 1/{d_E^{\text{eff}}}$.

The partial trace act on the entire global space $ \tr_E:\mathcal{L}(\mathcal{H}_S\otimes\mathcal{H}_E)\rightarrow\mathcal{L}(\mathcal{H}_S)$ and $\tr_S:\mathcal{L}(\mathcal{H}_S\otimes\mathcal{H}_E)\rightarrow\mathcal{L}(\mathcal{H}_E)$. However, note that the microcanonical state $\mathds{1}_R/d_R$ is defined in the restricted space $\mathcal{H}_R\subseteq\mathcal{H}_S\otimes\mathcal{H}_E$, so it is convenient we choose a basis $ \{\ket{R_r}\}\in\mathcal{H}_R$ in which elements can be written as:
\begin{align}
	\ket{R_r}=\sum_{ij}c^r_{ij}\ket{i}\otimes\ket{j}
\end{align}
with $ \ket{i} $ and $ \ket{j} $ orthonormal basis for $ \mathcal{H}_S $ and $ \mathcal{H}_E $, respectively.

In order to find the Choi state, let us first consider the maximally entangled state $ \ket{phi_+}\in\mathcal{H}_R\otimes\mathcal{H}_R $, in which its density matrix can be conveniently written as:
\begin{align}
	\ket{\phi_+}\bra{\phi_+}&=\dfrac{1}{d_R}\sum_{r,t}\ket{R_r}\bra{R_t}\otimes\ket{R_r}\bra{R_t} \nonumber \\
	&=\dfrac{1}{d_R}\sum_{r,t}\left(\sum_{i,j,k,l}c^r_{ij}c^{t\ast}_{k,l}\ket{ij}\bra{kl}\right)\otimes\ket{R_r}\bra{R_t} \nonumber \\ 
\end{align} 
So the Choi state related to the map $ \tr_{E} $ is given by
\begin{align}
	J_{\tr_{E}}&=(\tr_{E}\otimes\mathds{1}_R)(\ket{\phi_+}\bra{\phi_+})=\dfrac{1}{d_R}\sum_{r,t}\left(\sum_{i,j,k,l}c^r_{ij}c^{t\ast}_{k,l}\tr_E(\ket{ij}\bra{kl})\right)\otimes\ket{R_r}\bra{R_t} \nonumber \\ 
	&=\dfrac{1}{d_R}\sum_{i,k,r,t}\left(\sum_{j}c^r_{ij}c^{t\ast}_{k,j}\right)\ket{i}\bra{k}\otimes\ket{R_r}\bra{R_t}
\end{align} 
Calculating the purity
\begin{align}
	\tr(J_{\tr_{E}}^2)&=\tr\left[\dfrac{1}{d_R^2}\sum_{i,k,r,t,m,n,p,q}\left(\left(\sum_{j}c^r_{ij}c^{t\ast}_{kj}\right)\ket{i}\bra{k}\otimes\ket{R_r}\bra{R_t}\right)\left(\left(\sum_{z}c^p_{mz}c^{q\ast}_{nz}\right)\ket{m}\bra{n}\otimes\ket{R_p}\bra{R_q}\right)\right] \nonumber \\
	&=\dfrac{1}{d_R^2}\sum_{i,k,r,t,m,n,p,q}\left(\sum_{j}c^r_{ij}c^{t\ast}_{kj}\right)\left(\sum_{z}c^p_{mz}c^{q\ast}_{nz}\right)\delta_{km}\delta_{tp}\delta_{in}\delta_{rq}  \nonumber \\
	&=\dfrac{1}{d_R^2}\sum_{j,z}\left(\sum_{j}c^r_{ij}c^{t\ast}_{kj}\right)\left(\sum_{z}c^t_{kz}c^{r\ast}_{iz}\right)=\dfrac{1}{d_R^2}\sum_{j,z}\left(\sum_{r,i}c^r_{ij}c^{r\ast}_{iz}\right)\left(\sum_{t,k}c^t_{kz}c^{t\ast}_{kj}\right) 
	\label{eq:puritychoitre}
\end{align}

Since we want to prove $ \tr(J_{\tr_{E}}^2)=1/d_E^\text{eff}$ and the effective dimension of the environment is defined as $ d_E^\text{eff} = \tr(\Omega_{\tr_S}^2)$, let us explicitly describe the state $ \Omega_{\tr_S} $:
\begin{align}
	\Omega_{\tr_S} =& \tr_S\left(\dfrac{\mathds{1}_R}{d_R}\right)=\dfrac{1}{d_R}\tr_S\left(\sum_r\ket{R_r}\bra{R_r}\right) \nonumber\\
	&=\dfrac{1}{d_R}\tr_S\left(\sum_{r}\sum_{i,j,k,l}c^r_{ij}c^{r\ast}_{kl}\ket{ij}\bra{kl}\right)\nonumber\\
	&=\dfrac{1}{d_R}\sum_{j,l}\left(\sum_{r,i}c^r_{ij}c^{r\ast}_{il}\right)\ket{j}\bra{l}
\end{align}  
Now, lt us calculate the purity,
\begin{align}
	\tr\left(\Omega_{\tr_S}^2\right)&=\tr\left[\dfrac{1}{d_R^2}\sum_{j,l,m,n}\left(\left(\sum_{r,i}c^r_{ij}c^{r\ast}_{il}\right)\ket{j}\bra{l}\right)\left(\left(\sum_{t,k}c^t_{km}c^{t\ast}_{kn}\right)\ket{m}\bra{n}\right)\right] \nonumber \\
	&=\dfrac{1}{d_R^2}\sum_{j,l,m,n}\left(\sum_{r,i}c^r_{ij}c^{r\ast}_{il}\right)\left(\sum_{t,k}c^t_{km}c^{t\ast}_{kn}\right)\delta_{l,m}\delta_{j,n} \nonumber \\
	&=\dfrac{1}{d_R^2}\sum_{j,l}\left(\sum_{r,i}c^r_{ij}c^{r\ast}_{il}\right)\left(\sum_{t,k}c^t_{kl}c^{t\ast}_{kj}\right)
\end{align}
We can easily check that the above equation is equal to (\ref{eq:puritychoitre}) if we change the indices labels $ l\rightarrow z $, so we finish the demonstration.

\section{IV. Blurred and Saturated channel $ \Lambda_\text{BnS}^{n\rightarrow1} $ and derivation of $ \Omega_\Lambda^{\Lambda_\text{BnS}^{N\rightarrow k}} $} 

Since we split the $N$ atoms of the total system into $k = \frac{N}{n}$ blocks of $n$ atoms, the effective system will be equivalent to $k$ two-level atoms. Our starting point is the expression
\begin{equation}
	\Omega_{\Lambda_{\textrm{BnS}}} = \frac{1}{{N\choose Np}} \sum_{s \in \{0,1\}^k: |s| \in \{\ceil{k p}, ..., \min(N p, k)\}} \sum_{n_1 + n_2 + ... + n_{|s|} = Np - |s|} {\frac{N}{k}\choose n_1 + 1} {\frac{N}{k}\choose n_2 + 1} ... {\frac{N}{k}\choose n_{|s|} + 1} \proj{s}.
\end{equation}
In the expression above, $|s|$ is the number of 1's in the string $s$ and $n_{i} < \frac{N}{k}$, $\forall i \in [\, |s|\, ]$. Simplifying the second summations, 
\begin{equation}
	S = \sum_{n_1 + n_2 + ... n_{|s|} = Np - |s|}  {\frac{N}{k}\choose n_1 + 1} {\frac{N}{k}\choose n_2 + 1} ... {\frac{N}{k}\choose n_{|s|} + 1},
\end{equation}
with $n_{i} < \frac{N}{k}$. Therefore, $n_{i} \leqslant \frac{N}{k} - 1$ and
\begin{equation}
	S = \sum_{n_1 = 0}^{\frac{N}{k} - 1} \sum_{n_2 = 0}^{\frac{N}{k} - 1} ... \sum_{n_{|s|} = 0}^{\frac{N}{k} - 1} {\frac{N}{k}\choose n_1 + 1} {\frac{N}{k}\choose n_2 + 1} ... {\frac{N}{k}\choose n_{|s|} + 1} \delta(n_1 + n_2 + ... + n_{|s|}-( Np - |s|)).
\end{equation}

Noticing that 
\begin{equation}
	\delta(n_1 + n_2 + ... + n_{|s|} -( Np - |s|)) = \frac{1}{2 \pi} \int_{0}^{2 \pi} d\theta \,\, e^{i \theta n_1 + n_2 + ... + n_{|s|} - Np - |s|},
\end{equation}
\begin{equation}
	S = \frac{1}{2 \pi} \int_{0}^{2 \pi} d\theta \,\, e^{-i \theta (Np - |s|)} \bigg(\sum_{n=0}^{\frac{N}{k}-1} {\frac{N}{k} \choose n+1} e^{i \theta n}\bigg)^{|s|}.
\end{equation}
Making $m = n+1$, 
\begin{equation}
	\begin{split}
		S &= \frac{1}{2 \pi} \int_{0}^{2 \pi} d\theta \,\, e^{-i \theta (Np - |s|)} \bigg(\sum_{m=1}^{\frac{N}{k}} {\frac{N}{k} \choose m} e^{i \theta (m-1)}\bigg)^{|s|} \\
		&= \frac{1}{2 \pi} \int_{0}^{2 \pi} d\theta \,\, e^{-i \theta (Np - |s|)} \bigg(\sum_{m=0}^{\frac{N}{k}} {\frac{N}{k} \choose m} e^{i \theta (m-1)} - e^{-i \theta}\bigg)^{|s|} \\
		&=  \frac{1}{2 \pi} \int_{0}^{2 \pi} d\theta \,\, e^{-i \theta Np} \bigg(\sum_{m=0}^{\frac{N}{k}} {\frac{N}{k} \choose m} e^{i \theta m} - 1\bigg)^{|s|}. \\
		\Rightarrow S &=  \frac{1}{2 \pi} \int_{0}^{2 \pi} d\theta \,\, e^{-i \theta Np} \sum_{q=0}^{|s|} {|s|\choose q} \bigg[\sum_{m=0}^{\frac{N}{k}} {\frac{N}{k} \choose m} e^{i \theta m}\bigg]^{q} (-1)^{|s|-q}.
	\end{split}
\end{equation}

Therefore, 
\begin{equation}
	\begin{split}
		S &= \sum_{q=0}^{|s|} {|s| \choose q} (-1)^{|s| - q} \frac{1}{2 \pi} \int_{0}^{2 \pi} d\theta e^{-i \theta Np} \sum_{m_1 = 0}^{\frac{N}{k}} {\frac{N}{k} \choose m_1} e^{i \theta m_1} ... \sum_{m_q = 0}^{\frac{N}{k}} {\frac{N}{k} \choose m_q} e^{i \theta m+q} \\
		&= \sum_{q=0}^{|s|} {|s| \choose q} (-1)^{|s| - q} \sum_{m_1 = 0}^{\frac{N}{k}} \sum_{m_2 = 0}^{\frac{N}{k}} ... \sum_{m_q = 0}^{\frac{N}{k}} {\frac{N}{k} \choose m_1} {\frac{N}{k} \choose m_2} ... {\frac{N}{k} \choose m_q} \delta(m_1 + m_2 + ... + m_q - Np) \\
		&= \sum_{q=0}^{|s|} {|s| \choose q} (-1)^{|s| - q} \sum_{m_1 + m_2 + ... + m_q = Np}  {\frac{N}{k} \choose m_1}  {\frac{N}{k} \choose m_2} ...  {\frac{N}{k} \choose m_q} \\
		\Rightarrow S &=  \sum_{q=0}^{|s|} {|s| \choose q} (-1)^{|s| - q} {\frac{Nq}{k} \choose Np}.
	\end{split}
\end{equation}

The canonical state is given by 
\begin{equation}
	\Omega_{\Lambda_{\textrm{BnS}}} = \frac{1}{d_R} \sum_{s \in \{0,1\}^k: |s| \in \{\ceil{k p}, ..., \min(N p, k)\}} \sum_{q=0}^{|s|} {|s| \choose q}{\frac{Nq}{k} \choose Np} (-1)^{|s|-q} \,\, \Pi_{|s|},
\end{equation}
with $d_R = {N \choose Np}$ and $\Pi_{|s|}$ the projector onto the subspace spanned by the strings with number os 1's equal $|s|$. Finally, 
\begin{equation}
	\Omega_{\Lambda_{\textrm{BnS}}} = \frac{1}{d_R} \sum_{|s| = \ceil{k p}}^{\min(N p, k)} \sum_{q=0}^{|s|} {|s| \choose q}{\frac{Nq}{k} \choose Np} (-1)^{|s|-q} \,\, \Pi_{|s|}.
\end{equation}

\section{V. Linear entropy of $d_R$-depolarizing channel: $ S_L(\Lambda_{d_R}^\lambda) $}

The $d_R$-depolarizing channel $\Lambda_{d_R}^\lambda:\mathcal{L}(\mathcal{H}_R)\rightarrow\mathcal{L}(\mathcal{H}_R)$ is given by
\begin{align}
	\Lambda_{d_R}^\lambda(O)=\lambda \tr({O}) \dfrac{\mathds{1}_R}{d_R} + (1-\lambda)O,
\end{align}
with $O$ a linear operator acting on $\mathcal{H}_R$, and $0\leq \lambda \leq 1 + 1/(d_R^2-1)^2$.

Let us start by the Choi's state
\begin{align}
	J_{\Lambda_{d_R}^\lambda}&=\Lambda_{d_R}^\lambda\otimes\mathds{1}_R(\phi_+) \nonumber \\
	&=\Lambda_{d_R}^\lambda\otimes\mathds{1}_R\bigg(\dfrac{1}{d_R}\sum_{n,m}\ket{n}\bra{m}\otimes\ket{n}\bra{m}\bigg) \nonumber \\
	&= \dfrac{1}{d_R}\sum_{n,m}\Lambda_{d_R}^\lambda(\ket{n}\bra{m})\otimes\ket{n}\bra{m}   \nonumber\\
	&=\dfrac{1}{d_R}\sum_{n,m}\bigg[\lambda \underbrace{\tr(\ket{n}\bra{m})}_{\delta_{n,m}} \dfrac{\mathds{1}_R}{d_R} + (1-\lambda)\ket{n}\bra{m}\bigg]\otimes\ket{n}\bra{m}   \nonumber \\
	&= \dfrac{1}{d_R}\sum_{n}\left[\lambda\dfrac{\mathds{1}_R}{d_R} + (1-\lambda)\ket{n}\bra{n}\right]\otimes\ket{n}\bra{n} + \dfrac{1}{d_R}\sum_{n\neq m}\big[(1-\lambda)\ket{n}\bra{m}\big]\otimes\ket{n}\bra{m} \\
	&= \lambda\dfrac{\mathds{1}_R}{d_R}\otimes\bigg(\dfrac{1}{d_R}\underbrace{\sum_{n}\ket{n}\bra{n}}_{\mathds{1}_R}\bigg) +(1-\lambda)\dfrac{1}{d_R}\sum_{n} \ket{n}\bra{n}\otimes\ket{n}\bra{n}+ (1-\lambda)\dfrac{1}{d_R}\sum_{n\neq m}\ket{n}\bra{m}\otimes\ket{n}\bra{m} \nonumber \\
	&=\lambda\dfrac{\mathds{1}_R}{d_R}\otimes\dfrac{\mathds{1}_R}{d_R}+(1-\lambda)\phi_+.	
\end{align}

Since the linear entropy is defined as $S_L(\Lambda)=1-\operatorname{tr}(J_\Lambda^2)$, let's now compute $J_{\Lambda_{d_R}^\lambda}^2$

\begin{align}
	J_{\Lambda_{d_R}^\lambda}^2&=\left[\lambda\dfrac{\mathds{1}_R}{d_R}\otimes\dfrac{\mathds{1}_R}{d_R}+(1-\lambda)\phi_+\right]^2 \nonumber \\
	&=\dfrac{\lambda^2}{d_R^4}\mathds{1}_R\otimes\mathds{1}_R+(1-\lambda)^2\phi_+ +\dfrac{2\lambda(1-\lambda)}{d_R^2}\phi_+. 
\end{align}
Evaluating the trace
\begin{align}
	\tr\left[J_{\Lambda_{d_R}^\lambda}^2\right]&=\dfrac{\lambda^2}{d_R^4}\tr\left[\mathds{1}_R\otimes\mathds{1}_R\right]+(1-\lambda)^2 \tr \left[\phi_+\right] +\dfrac{2\lambda(1-\lambda)}{d_R^2}\tr\left[\phi_+\right] \nonumber \\
	&=\dfrac{\lambda^2}{d_R^2}+(1-\lambda)^2+\dfrac{2\lambda(1-\lambda)}{d_R^2} \nonumber \\
	&=\dfrac{\lambda(2-\lambda)}{d_R^2}+(1-\lambda)^2.
\end{align}

Finally, the linear entropy of the $d_R$-depolarizing channel is
\begin{align}
	S_L(\Lambda_{d_R}^\lambda)&=1-\tr\left[J_{\Lambda_{d_R}^\lambda}^2\right] \nonumber \\
	&=1-\dfrac{\lambda(2-\lambda)}{d_R^2}-(1-\lambda)^2.
\end{align}

\end{document}